# Promoting and imaging intervalley coherent order in rhombohedral tetralayer graphene on MoS$_2$

Wei-Yu Liao[1,4,†], Wen-Xiao Wang[2,†], Shihao Zhang[1,†], Yang Zhang[1,4], Ling-Hui Tong[1,4], Wenjia Zhang[2], Hao Cai[1,4], Yuan Tian[1], Yuanyuan Hu[3], Li Zhang[1], Lijie Zhang[1], Zhihui Qin[1], and Long-Jing Yin[1,4,*]

[1] *Key Laboratory for Micro/Nano Optoelectronic Devices of Ministry of Education & Hunan Provincial Key Laboratory of Low-Dimensional Structural Physics and Devices, School of Physics and Electronics, Hunan University, Changsha 410082, China*

[2] *College of Physics and Hebei Advanced Thin Films Laboratory, Hebei Normal University, Shijiazhuang 050024, China*

[3] *College of Semiconductors (College of Integrated Circuits), Hunan University, Changsha 410082, China*

[4] *Research Institute of Hunan University in Chongqing, Chongqing 401120, China*

[†]These authors contributed equally to this work.

*Corresponding author: yinlj@hnu.edu.cn

**Multilayer rhombohedral graphene (RG) has recently emerged as a new, structurally simple flat-band system, which facilitates the exploration of interaction-driven correlation states with highly ordered electron arrangements. Despite a variety of many-body order behaviors observed in RG by transport measurements, the direct microscopic visualization of such correlated phases in real space is still lacking. Here, we show the discovery of a robust intervalley coherent order—a long-predicted ground state in RG—at 77 K in tetralayer RG placed on MoS$_2$ via imaging atomic-scale spatial reconstruction of wave functions for correlated states. By using scanning tunnelling microscopy, we observe spectroscopic signatures of electronic correlations at partially filled flat bands, where distinct splitting appears. At ~60% and ~70% fillings of the flat bands, we visualize atomic-scale reconstruction patterns with a $\sqrt{3}\times\sqrt{3}$ supercell on**

**graphene lattice at liquid nitrogen temperature, which indicates a robust intervalley coherent phase of the interacting electrons. The $\sqrt{3} \times \sqrt{3}$ pattern is observed in MoS$_2$-supported RG, while it is absent in hBN-based ones under the same experimental conditions, suggesting the significant influence of spin–orbit proximity effect. Our results provide microscopic insights into the correlated phases in tetralayer RG and highlight the significant potential for realizing highly accessible collective phenomena through Van der Waals proximity.**

In the presence of inter-particle interactions, the combined spin and valley degrees of freedom of electrons in graphene can give rise to various phases with spontaneously broken symmetry and highly ordered electron arrangements. Such symmetry-broken correlated states usually share nearly degenerate energies and tend to persist at extremely low temperatures, making it challenging to precisely determine their ground states and impeding their accessibility. Recently, the improvements in low-temperature scanning tunnelling microscopy and spectroscopy (STM and STS) have significantly enhanced the spatial imaging of electron wave functions, enabling access to crucial microscopic information regarding quantum phases[1-6]. This has facilitated the identification of competing broken symmetry states in several correlated graphene systems, such as magnetic-field-induced Landau levels[7-9] and twisting-created moiré flat bands[10,11], albeit achieved at extremely low temperatures.

Multilayer rhombohedral graphene (RG; or *ABC*-stacked graphene) has been recently demonstrated to be a structurally simple and strongly correlated system due to the existence of intrinsic low-energy flat bands. Transport experiments have discovered a variety of symmetry breaking phases[12-15] including ferromagnetism[16], ferro-valleytricity[17], and superconductivity[18] in RG. More recently, experimental signatures of unexpected robust collective phenomena, such as rich fractional quantum anomalous Hall states[19,20] and large integer quantum anomalous Hall effects[21,22], have been observed in RG system through the introduction of more graphene layers or heterostructure substrates[23-26]. This indicates the great potential for the manifestation of more accessible many-body physics in RG-based materials. However, the direct microscopic visualization of the correlated quantum states and their stability against thermal fluctuations in RG remains unexplored, which hampers our understanding of its multiple ground states and thus the underlying physics mechanisms.

Here we show a direct observation of interaction-driven intervalley coherent (IVC) order of electrons in tetralayer (4L) RG placed on MoS$_2$ substrate at unexpected liquid nitrogen temperature. This IVC phase has been extensively predicted to be one of the ground states with spontaneously broken symmetry in RG[27,28], which exhibits a highly enigmatic correlation with superconductivity but has yet to be directly observed. We determine the IVC order from the imaging of atomic-scale reconstruction patterns with a $\sqrt{3} \times \sqrt{3}$ supercell on graphene lattice through real space measurements of the local density of states (LDOS) in an STM. This order state is observed at ~60% and ~70% filling ratios of the 4L RG flat bands, where a clear splitting of the flat band is detected. Moreover, we find the absence of the IVC order in hBN-based 4L RG under the same experimental conditions, indicating a significant influence from the spin-orbit proximity effect. These observations are accurately reproduced by our theoretical calculations. Our experiments directly confirm the existence of the IVC order in RG, demonstrating that RG multilayers can host unexpected robust correlated quantum phases against thermal fluctuations, which could be further enhanced through Van der Waals proximity.

Our tetralayer RG samples were stacked on thick MoS$_2$ layers deposited on a SiO$_2$/Si substrate using the normal van der Waals stacking technique (Methods). The samples were measured by the STM and STS at 77 K. Figure 1a shows the schematic of our STM measurement for the 4L RG/MoS$_2$ device. Figure 1b,c shows a representative large-scale STM topographic image of a tetralayer RG/MoS$_2$ (device B) and its atomic-resolution image, respectively. The RG sample displays a relatively flat and clean surface. From a series of atomically resolved STM images, we obtained a homogeneous lattice constant of 0.246 ± 0.001 nm over three lattice directions, which aligns with the theoretical value of graphene and suggests the absence of detectable strain in the MoS$_2$-supported 4L RG. Similar topographic features were observed in all three of our devices (see Supplementary Figs. 1 and 2 for more data). Moreover, due to the large lattice mismatch δ between graphene and $MoS_2$ (δ ~ 0.3[29]), no substrate-induced moiré superlattice is expected for the $RG/MoS_2$, as evidenced by the topographic images of different sizes (Fig. 1b,c and Supplementary Fig. 1). The above results demonstrate that $MoS_2$ is a favorable substrate to support graphene layers[30]. Further evidence for $MoS_2$ being a suitable substrate can be found from the measurement of its electronic structures. We measured the differential conductance *dI/dV*—reflecting the electronic LDOS of the sample surface—at exposed $MoS_2$ region and obtained an energy gap of ~1.23 eV (Supplementary Fig. 3). This gap value is consistent with those reported in thick $MoS_2$[31-33]. The measured valence band edge and conduction band edge of the $MoS_2$ are located at ~-0.75 eV and ~0.5 eV, respectively. Over this energy gap range, most of the low-energy band structures of tetralayer RG, including the flat bands and the remote-band van Hove singularities (see below), can be effectively preserved, allowing access to its intrinsic low-energy electronic properties.

Figure 1d,e shows the spatially resolved *dI/dV* spectra and a representative point spectrum of the 4L RG, respectively (a typical V-shaped STS spectrum of *ABA*-stacked tetralayer region is also presented in Fig. 1e for comparison). The STS spectra exhibit a pronounced DOS peak near the Fermi energy, which originates from the zero-energy flat bands of the RG[34-39]. The flat-band DOS peak shows high spatial uniformity over hundreds of nanometers (Fig. 1d), further indicating the excellent quality of the sample. Away from the flat-band peak, the DOS initially remains relatively constant and then sharply increases at a certain high energy. This is because the DOS of RG has additional contributions from the remote bands at high energies (Fig. 1f). By examining various spatially resolved *dI/dV* spectra, we identified the positions of the remote-band edges at ~±0.3 eV (see Supplementary Fig. 1 for details). Based on these spectroscopic characteristics, we can reproduce the band structures of the tetralayer RG by using the tight-binding model with a Slonczewski-Weiss-McClure (SWMcC) parametrization. Among the SWMcC theoretical model, the nearest-neighbor interlayer coupling strength $\gamma_1$ is the dominated parameter that controls the locations of the remote-band

edges in RG[40,41]. We thus set $\gamma_1$ as the free parameter and utilize other coupling terms extracted from theory[42]. The calculated band structures and the corresponding LDOS, which best fit our experimental data, are shown in Fig. 1f, resulting in $\gamma_1 \approx 0.38$ eV. This obtained value of $\gamma_1$ is consistent with both the theoretical prediction[42] and the measurement in 4L RG on hBN[39]. We also used the most simplified SWMcC model (considering only $\gamma_0$ and $\gamma_1$) to fit our result and deduced a nearly identical $\gamma_1$, indicating a negligible influence of other weaker coupling parameters on the determination of $\gamma_1$ here (see Supplementary Fig. 4 for a detailed discussion).

We next investigate the electron correlations of the flat bands in the $MoS_2$-based 4L RG. As previously reported, the substrate can introduce charge inhomogeneity on the surface of supported graphene, leading to differently doped regions[43,44]. By moving the STM probe tip on a micrometer scale, we can find differently doped *ABC* regions in our $MoS_2$-supported 4L RG, similar to that observed in hBN-based graphene[39]. We identify doping-varied locations through the position shift of the flat-band peak (i.e., charge neutrality point, CNP). Figure 2a shows *dI/dV* spectra of the 4L RG recorded in three differently doped regions, where the flat-band peak exhibits fully filled, partially filled, and empty states respectively. These filling-varied STS spectra enable a direct exploration of electron correlations within the flat bands[44]. Obviously, the flat-band peak broadens in width and splits into two DOS peaks at partial fillings (Fig. 2b,c), while maintaining a narrower single peak at full-filled and unfilled states. This partial filling-induced broadening and splitting of flat bands were observed in many positions of the RG sample with various partial-filling states. It is worth noting that for each differently doped region, the STS spectra all display a local uniformity (Fig. 1d and Supplementary Fig. 1). The local charge potential fluctuation induced scattering and confinement effects can therefore be excluded here. The above spectroscopic phenomena demonstrate explicitly the emergence of interaction-driven strongly correlated states, as observed in RG on other substrates[39,44,45] and in magic-angle twisted graphene[1-3].

To access more information on electron correlations, we examined the splitting energies of the flat bands under partial-filling states. The splitting energy of the flat-band peak is extracted by measuring peak-to-peak separation between two Gaussian fitting peaks, as exemplified in Fig. 2b. We obtained a splitting energy of 51 ± 9 meV for the partially filled flat bands in our 4L RG on $MoS_2$. This value is consistent with the interaction-induced spectroscopic splitting of flat bands recently observed in 4L RG placed on hBN at the same temperature of 77 K[39], indicating the prevalence of strong correlations in RG systems.

We now focus on investigating the influence of electron interactions on the electronic states of RG in real space. Figure 3a,b shows the atomic-resolution STM topographic images measured in the same RG region (device A), with the flat-band peak being approximately 70% filled and notably split (Fig. 3e), at different bias voltages. Here, the filling ratio is defined as the ratio between the area of the flat-band *dI/dV* peaks below the Fermi level and the total area of the peaks, similar to that used in magic-angle bilayer graphene[1] and rhombohedral graphite[44] (see Supplementary Fig. 5 for details). At 300 mV, the STM image exhibits a characteristic triangular lattice structure typical of multilayer graphene, with a lattice constant of ~0.246 nm. However, at the reduced bias of 40 mV, an atomic-scale modulation that triples the unit cell size of graphene is observed, indicating symmetry breaking on the graphene lattice scale. Such tripling-lattice modulated pattern is further confirmed by the fast Fourier transform (FFT) analysis of the STM images. As shown in Fig. 3c,d, in addition to the six outer spots corresponding to the graphene lattice, there are six inner spots displayed at positions that correspond to $1/\sqrt{3}$ of the graphene reciprocal lattice vectors rotated by 30° for the low-energy FFT image. The observed $\sqrt{3} \times \sqrt{3}$ supercell pattern demonstrates the existence of the IVC order arising from a coherent superposition of electron wavefunctions between the *K* and *K'* valleys, i.e., the Kekulé distortion[27,28].

We further examine the dependence of the IVC pattern on bias voltage and filling state. Figure 3f,g displays a series of *dI/dV* mappings at different bias voltages ranging from -500 mV to 500 mV (Fig. 3f), along with their corresponding FFT images (Fig. 3g). The remarkable $\sqrt{3} \times \sqrt{3}$ superlattice structure depends sensitively on the applied bias: it only manifests at biases aligning with the flat-band peak. This result can be further supported by the extracted IVC intensity normalized by the intensity of the graphene lattice from the FFT images (Fig. 3h). Notably, the non-zero IVC intensity only exists in the bias region matching the flat-band peak, and exhibits its maximum near Fermi level—where the 2-order IVC even can be seen clearly. In addition, the IVC pattern also displays a strong filling correlation. Figure 4a shows low-bias atomic-resolution STM images measured in differently filled RG regions of the same device (device A). The $\sqrt{3} \times \sqrt{3}$ reconstruction structure is observed in ~60% and ~70% filled flat bands, while it is absent at the RG regions with ~50% (i.e., the CNP) and ~100% filling ratios (Fig. 4b,c and see Supplementary Figs. 6 and 7 for more data). The above findings, which include the strong bias and filling dependence of the $\sqrt{3} \times \sqrt{3}$ superlattice, along with the defect- and strain-free local topography, essentially rule out the scattering mechanism induced by disorder and explicitly indicate an electronic interaction origin for the observed IVC pattern.

To investigate the underlying mechanism of this unexpected phenomenon, we performed similar atomic-scale measurements of electronic states in 4L RG placed on hBN substrates. No atomic-scale reconstruction signature has been detected in the hBN-supported RG either for partial filling or non-partial filling states under the same experimental conditions (see Supplementary Figs. 8-10 for details). This suggests that the spin-orbit coupling (SOC) effect, proximitized by the $MoS_2$ substrate, may play a crucial role in facilitating the robust IVC order in RG.

Symmetry-broken induced IVC reconstruction phases, which exhibits highly similar microscopic characteristics to ours, have been recently observed in strongly correlated regions of flat bands in magic-angle twisted graphene bilayer and trilayer under liquid helium temperature[10,11]. For RG system, theory has predicted the IVC ground state with isospin symmetry breaking in partially filled flat band of crystalline trilayer[27,28]. Very recently, switchable isospin induced IVC phase has been proposed through the introduction of proximity-induced SOC from transition metal dichalcogenide substrates in RG multilayers[46-48]. However, a direct experimental confirmation of such IVC state is still lacking[49]. Our microscopic experiment explicitly demonstrates the existence of the IVC order in RG, with the observed filling ranges aligning well with theoretical predictions (i.e., a finite doping region). More surprisingly, the robust IVC order in the MoS$_2$-based 4L RG are observed under a liquid nitrogen environment.

To support our observations and further understand the underlying physics, we carried out Hartree-Fock mean-field calculations. Our Hartree-Fock calculations find two types of IVC states in the MoS$_2$-supported tetralayer RG, which we refer to as IVC$_0$ and SVL-IVC (Fig. 5). Firstly, we discuss the noninteracting states. The proximity-induced Ising SOC results in a spin-valley-locked (SVL) state in 4L RG adjacent to MoS$_2$ (Fig. 5a). Its order parameter can be described as $\tau_z s_z$, where $\tau$ and $s$ denote the Pauli matrices in the valley and spin spaces, respectively, as presented on the Bloch sphere in Fig. 5c. In the SVL state, the spin rotation symmetry $SU(2)_s$ is broken due to SOC effect. When electron interaction is taken into consideration, a IVC$_0$ state occurs in the 4L RG. The IVC$_0$ state is the mixture of IVC and SVL orders (Fig. 5d). The order parameter is summarized as $(\tau_x s_0, \tau_z s_z)$, which is consistent with previous theoretical results[48]. The spin rotation $SU(2)_s$ and valley $U(1)_v$ symmetries are both broken in this state, but the spin $U(1)_s$ symmetry is still preserved. However, when Ising SOC $\lambda_I$ is enhanced, the valley polarizations in each spin channel are also enhanced by Ising SOC, which will kill the IVC order under large Ising SOC. Thus, the IVC$_0$ state is unstable under large Ising SOC effect. Another IVC state, SVL-IVC, is shown in Fig. 5b,e, whose order parameter is $(\tau_x s_x, \tau_z s_z)$[46,47]. In this flavor-coherent state, the spin rotation $SU(2)_s$, valley $U(1)_v$, and spin $U(1)_s$ symmetries are all broken. Different from the IVC$_0$ state, the $\tau_x s_x$ flavor-coherent component can persist in the giant Ising SOC. The SVL-IVC state holds the effective time reversal symmetry $\mathcal{T} = \tau_y s_y \mathcal{K}$. It is a spatially modulated phase with the wave vector $\boldsymbol{K_0} = (0, 4\pi/3a)$ connecting two different valleys where $a$ is lattice constant of graphene (see Methods for more details). The spatially modulated spin polarization at the $\boldsymbol{r}$ point is proportional to $\cos(2\boldsymbol{K_0} \cdot \boldsymbol{r})$[46]. Thus, the SVL-IVC order holds the $\sqrt{3} \times \sqrt{3}$ supercell with spin-density wave phase, which is consistent with our observations.

It is worth noting that besides inducing the Ising SOC effect in the tetralayer RG, the MoS$_2$ substrate also leads to the stronger screening effect due to the virtual excitations of electron-hole pair. Pervious theoretical analysis shows that, compared to other flavor polarized states, the Fock-type energy penalty of the IVC state is proportional to the strength of electron interactions[27]. Here the strong screening effect from $MoS_2$ substrate weakens the electron-electron interactions, which assists the emergence of the IVC order in the 4L RG. Therefore, we speculate that the observed robust IVC order is closely associated with the cooperation between the SOC proximity effect and the strong screening effect from $MoS_2$.

Our microscopic experiments directly confirm the existence of the IVC phase in multilayer RG system. In particular, this correlated ordered phase was found to exist at liquid nitrogen temperature. Our results align well with the theoretical calculations, indicating that the thick $MoS_2$ substrate-induced Ising SOC effect, in conjunction with the finite screening, plays a crucial role in the emergence of the strong IVC order. These findings may prove valuable to understand the microscopic origin of superconductivity in RG systems[50-52]. Moreover, the presented results establish multilayer RG-based Van der Waals heterostructures as an ideal physics platform, opening up the possibility of probing diverse and robust collective phenomena, especially for the valley-related states, in the presence of thermal fluctuations. Further investigations, such as gate-tunable measurements and substrate-varied experiments, have the potential to unveil even richer phase diagrams in the proximity-modulated multilayer RG.

# Methods

## Sample preparations

The RG 4-layer and $MoS_2$ flake (80-120 nm thick) were exfoliated from natural bulk crystals and placed onto silicon substrates covered with a 285 nm thick oxide layer. Raman spectroscopy with a laser excitation of 488 nm and atomic force microscope were used to determine the *ABC* stacking order and to measure the sample thickness, respectively. The RG/$MoS_2$ devices were fabricated by the typical van der Waals stacking technique (Supplementary Fig. 11). We first annealed the $MoS_2$ flake at 350 °C for 4h in a hydrogen-argon mixture (1:10) and then stacked the RG layers onto the $MoS_2$ flake placed on $SiO_2$/Si. The RG flake was lifted by a handle composed of a PVA (Polyvinyl alcohol) film and a PDMS (Polydimethylsiloxane) stamp on a glass slide. We dissolved the PVA film in deionized water for an hour and then removed it. Au electrodes (50 nm thick) were evaporated onto the graphene layers using a stencil mask for electrical contact with the STM. Before transferring the devices to the STM chamber, we rechecked the *ABC* stacking order using Raman spectroscopy (Supplementary Figs. 12-14).

## STM/STS measurements

STM and STS measurements were performed in an ultrahigh vacuum CreaTec STM system using an electrochemically etched W tip. The STM system was operated in constant-current mode at liquid nitrogen temperature. The STM topographic images were calibrated on a standard Au(111) surface and a Si(111)-(7×7) lattice. The STS measurements were conducted using the standard lock-in technique, with an a.c. modulation of 793 Hz and 10-20 mV being added to the bias voltage. The *dI/dV* maps were measured under the constant-height mode by turning off the feedback circuit.

## The SOC effect in the tetralayer RG on $MoS_2$ substrate

When the tetralayer RG is set on the $MoS_2$ substrate, the CNP of graphene is lying in the gap of $MoS_2$. But the virtual electrons' tunneling between graphene and $MoS_2$ leads to the proximitized SOC in the graphene. There are three types of SOC effects in the graphene on the MoS2 substrate including Ising SOC $H_{\text{Ising}}$, Rashba SOC $H_{\text{Rashba}}$ and Kane-Mele SOC $H_{\text{KM}}$ terms[46-48,53-55].

$$H_{\text{Ising}} = \lambda_I \tau_z s_z \sigma_0,$$

$$H_{\text{Rashba}} = \lambda_R (\tau_z \sigma_x s_y - \sigma_y s_x),$$

$$H_{\text{KM}} = \lambda_{KM} \tau_z s_z \sigma_z.$$

Here $\lambda_I$, $\lambda_R$ and $\lambda_{KM}$ refer to Ising SOC, Rashba SOC and Ising SOC, respectively. And $\tau$, $s$ and $\sigma$ represent the Pauli matrices in the valley, spin and sublattice spaces, respectively. Kane-Mele SOC term is negligible compared to Ising SOC and Rashba SOC. Rashba SOC is off-diagonal in the sublattice subspace, so it has weak effect on the "flat-band" states with strong sublattice polarization in the tetralayer RG. Thus, we only consider the Ising SOC effect in our discussions about correlated electronic states.

**The Kekulé distortion in the SVL-IVC state**

The IVC order may lead to the Kekulé electronic state, but it is noted that the Kramers IVC (KIVC) state in the magic twisted bilayer graphene has no Kekulé distortion in real space[56]. In this section, we argue that the SVL-IVC state owns the Kekulé electronic state.

Now we consider the Fourier-transformed LDOS following previous work[27],

$$\rho(\boldsymbol{q}, E) = -\frac{1}{2\pi i}\left[\text{Tr}\left[\hat{\rho}(\boldsymbol{q})\hat{G}(E)\right] - \overline{\text{Tr}\left[\hat{\rho}(-\boldsymbol{q})\hat{G}(E)\right]}\right]$$

Here $\hat{\rho}(\boldsymbol{q}) = e^{iqr}$ is the density operator, and $\hat{G}(E)$ is the electronic Green function. The $\hat{\rho}(\boldsymbol{q} = \boldsymbol{K} - \boldsymbol{K}' + \Delta \boldsymbol{q})$ at the inter-valley momentum can contribute to the Kekulé LDOS.

In the KIVC state, the intervalley component is $\hat{\rho}(\boldsymbol{q} = \boldsymbol{K} - \boldsymbol{K}' + \Delta \boldsymbol{q}) = \sum_k |k + \Delta q, \tau\rangle\langle k, -\tau|$ in which $\boldsymbol{k}$ is the momentum near the Dirac point. The KIVC state hold the effective time reversal symmetry $\mathcal{T}'|k,\tau\rangle = \tau|-k,-\tau\rangle$, so $\mathcal{T}'^{-1}\hat{\rho}(\boldsymbol{q})\mathcal{T}' = -\hat{\rho}(-\boldsymbol{q})$ and Kekulé distortion disappears in the KIVC state[56].

But in the SVL-IVC state in the tetralayer RG, the intervalley component becomes $\hat{\rho}(\boldsymbol{q} = \boldsymbol{K} - \boldsymbol{K}' + \Delta\boldsymbol{q}) = \sum_k |\boldsymbol{k} + \Delta\boldsymbol{q}, \tau, s\rangle\langle \boldsymbol{k}, -\tau, -s|$. The effective time reversal symmetry in the SVL-IVC state keeps that $\mathcal{T}'|k, \tau, s\rangle = \tau s|-k, -\tau, -s\rangle$, so $\mathcal{T}'^{-1}\hat{\rho}(\boldsymbol{q})\mathcal{T}' = \hat{\rho}(-\boldsymbol{q})$ and non-zero real-space Kekulé distortion is allowed in the SVL-IVC state.

**The Hartree-Fock mean-field calculation method**

Our mean-field calculations adopt the dual-gate screened interaction potential $V_q = \frac{2\pi k_e}{\epsilon_r}\frac{\tanh(|q|d)}{|q|}$. Here $d$ is set to 120 nm and $k_e = 1.44$ eV nm is the Coulomb constant. We set the dielectric constant $\epsilon_r$ as 5.0 to consider the strong screening effect induced by $MoS_2$ substrate. To investigate the screening effect from virtual excitations of particle-hole pairs, we use the constraint random-phase approximation (cRPA) in the tetralayer graphene. The interaction potential becomes $V_q^{RPA} = \frac{V_q}{1+\chi V_q}$ in which $\chi$ is the susceptibility obtained by cRPA calculations. We solve the interacting Hamiltonian by Hartree-Fock approximation with the 96×96 moment grid.

**Acknowledgements**

This work was supported by the National Natural Science Foundation of China (Grant Nos. 12474166, 12174095, 12304217, 12174096, 62101185, 12204164 and 51972106), and the Natural Science Foundation of Hunan Province, China (Grant No. 2021JJ20026). L.-J.Y. also acknowledges support from the Science and Technology Innovation Program of Hunan Province, China (Grant No. 2021RC3037) and the Natural Science Foundation of Chongqing, China (cstc2021jcyj-msxmX0381). W.-X.W acknowledges the support from the Science Research Project of Hebei Education Department (BJK2024168). The authors acknowledge the financial support from the Fundamental Research Funds for the Central Universities of China.

**Author contributions**

L.-J.Y. designed and supervised the project. W.-Y.L. fabricated the devices and performed the STM measurements. Y.Z., L.-H.T., W.Z. and H.C. assisted with the STM measurements. L.-H.T., Y.T., Y.H. L.Z., L.Z. and Z.Q. assisted with the device fabrications. W.-Y.L. and L.-J.Y. analysed the data. W.-X.W. and L.-J.Y. supervised the STM measurements. S.Z. performed the mean-field calculations. W.-Y.L., S.Z. and L.-J.Y. wrote the manuscript with input from all the other authors.

**Data availability**

The data that support the findings of this work are available from the corresponding authors upon reasonable request.

**Code availability**

The code used for the modelling in this work are available from the corresponding authors upon reasonable request.

**Competing interests**

The authors declare no competing interests.

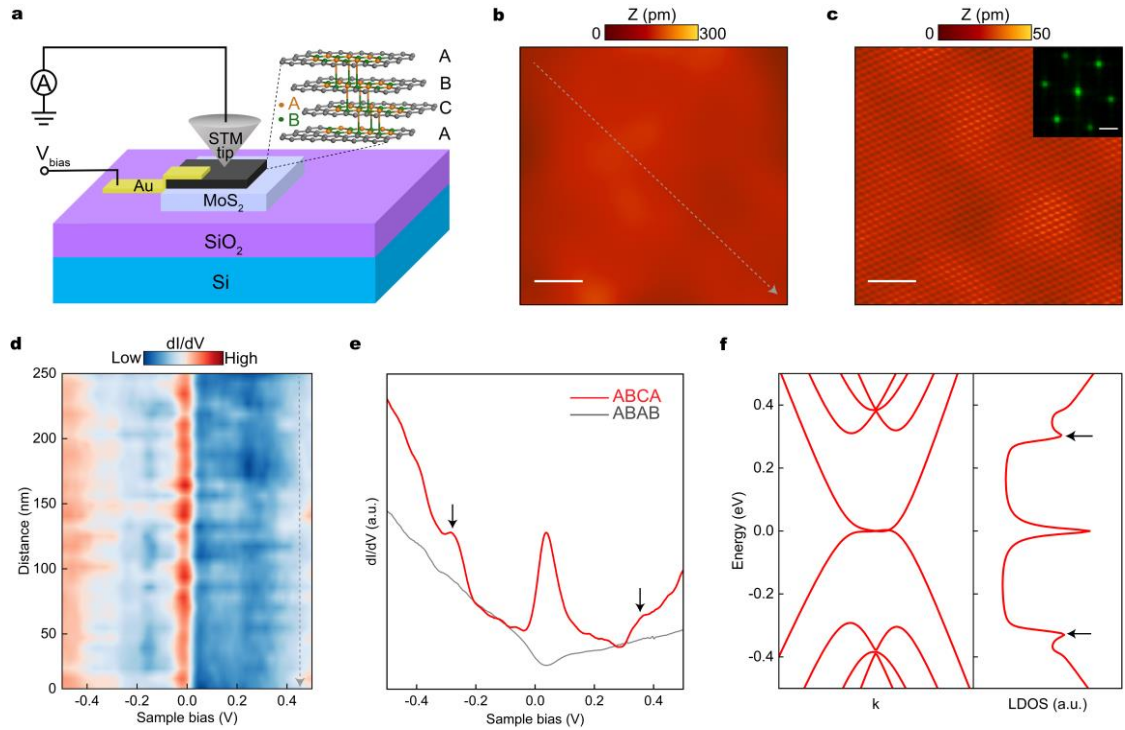

**Fig.1 | Topography and spectroscopy of rhombohedral 4-layer graphene. a,** Schematic diagram of the experimental setup. **b,** Large-scale STM topographic image (200 nm × 200 nm, $V_{bias}$ = -0.6 V, $I$ = 0.1 nA) of a 4-layer RG/MoS$_2$ heterostructure (device B). **c,** Atomic-resolution STM topographic image (10 nm × 10 nm, $V_{bias}$ = 0.3 V, $I$ = 0.1 nA) taken from **b**. Inset: Fourier transform of the STM image. **d,** Contour plot of spatially resolved $dI/dV$ spectra ($V_{bias}$ = -0.4 V, $I$ = 0.1 nA) measured in the ABC region of **b** at 77 K. **e,** Typical $dI/dV$ spectra ($V_{bias}$ = 0.4 V, $I$ = 0.1 nA) of the ABC and ABA 4L graphene measured in device B. **f,** Calculated band structures of tetralayer RG and the corresponding LDOS on the top layer obtained by the tight-binding model with hopping parameters: $\gamma_0$ = 3.16 eV, $\gamma_1$ = 0.384 eV, $\gamma_2$ = -0.0083 eV, $\gamma_3$ = 0.293 eV, $\gamma_4$ = 0.144 eV. Arrows in **e,f** mark the remote band edges. Scale bars: 40 nm (**b**), 2 nm (**c**), 2.5 nm$^{-1}$ (inset of **c**).

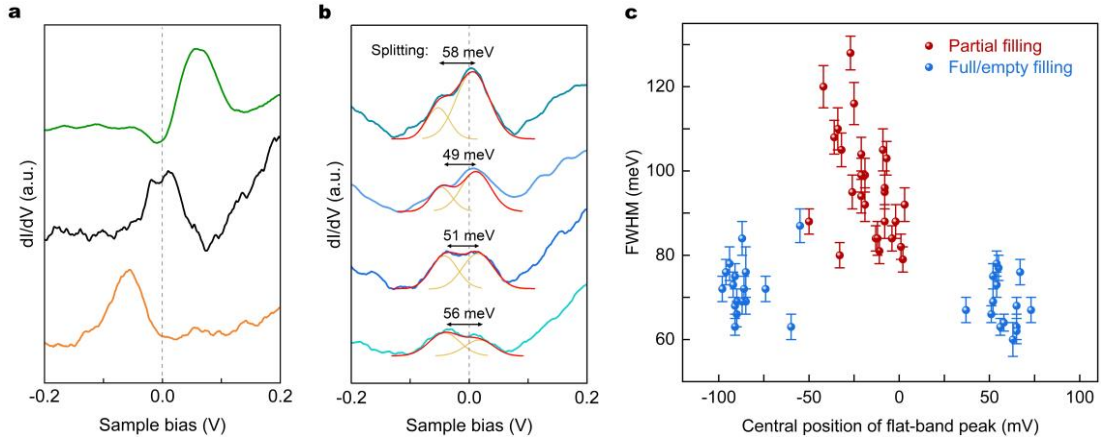

**Fig. 2 | Spectroscopy of electronic correlations in tetralayer RG. a,** Representative *dI/dV* spectra with different filling states of the flat-band peak (device A). **b,** Typical STS spectra of partially filled flat-band peak for device A. Double-Gaussian peak fits are shown by red curved lines for the flat-band splitting. The STS curves in **a** and **b** are shifted vertically for clarity. Set-point parameters: $V_{bias}$ = 0.6 V, $I$ = 0.1 nA. **c,** Full width at half maximum (FWHM) of the flat-band peak as a function of flat-band peak position. The FWHM was obtained from a single Gaussian fit to the flat-band peaks. The flat-band peak position was identified as the center position of the Gaussian fitting peak. Error bars were estimated by combining the fitting uncertainty and standard deviation of the data.

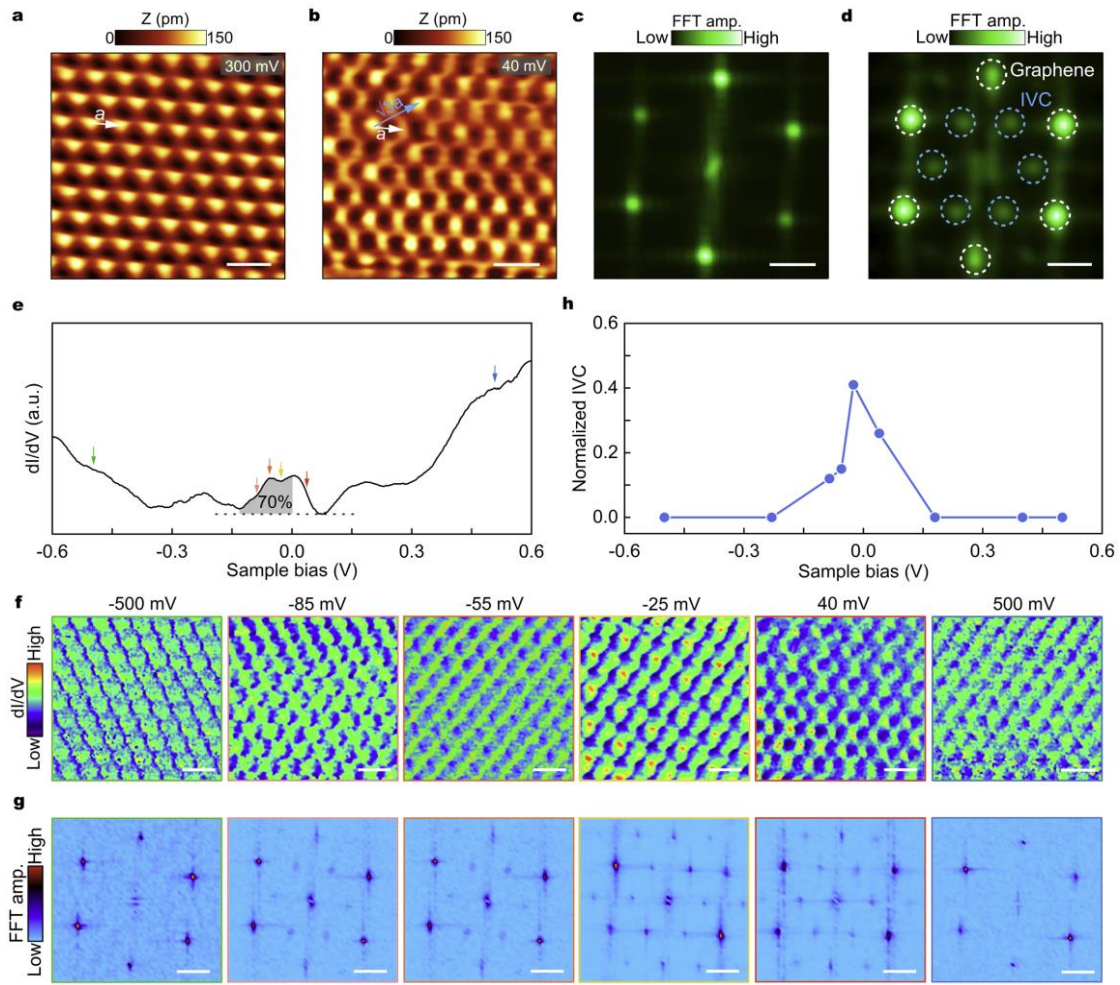

**Fig. 3 | Imaging atomic-scale IVC patterns**. **a,b,** Atomic-resolution STM topographic images (2 nm × 2 nm, $I$ = 0.1 nA) over the same region of the tetralayer RG (device A) for $V_{bias}$ = 300 mV (**a**) and 40 mV (**b**). **c,d,** FFT images of **a** and **b**, respectively. The black and blue dashed circulars mark the reciprocal lattice vectors of graphene lattice and IVC order. **e,** A typical $dI/dV$ spectrum for the region in **a** and **b**, showing nearly 70% filled flat-band peak. The colored arrows denote the bias parameters of the STS mapping in **f**. **f,** Atomic-resolution $dI/dV$ maps over the same region of **a** and **b** measured at different bias (from left to right: -500 mV, -85 mV, -55 mV, -25 mV, 40 mV, 500 mV). **g,** Corresponding FFT images of **f**, showing well-resolved spots for the graphene lattice and IVC order. **h,** Normalized IVC strength $\Sigma|FFT(IVC)/\Sigma FFT(graphene)|$ at different biases extracted from **g**. Scale bars: 4 Å (**a,b,f**); 2.5 nm$^{-1}$ (**c,d,g**).

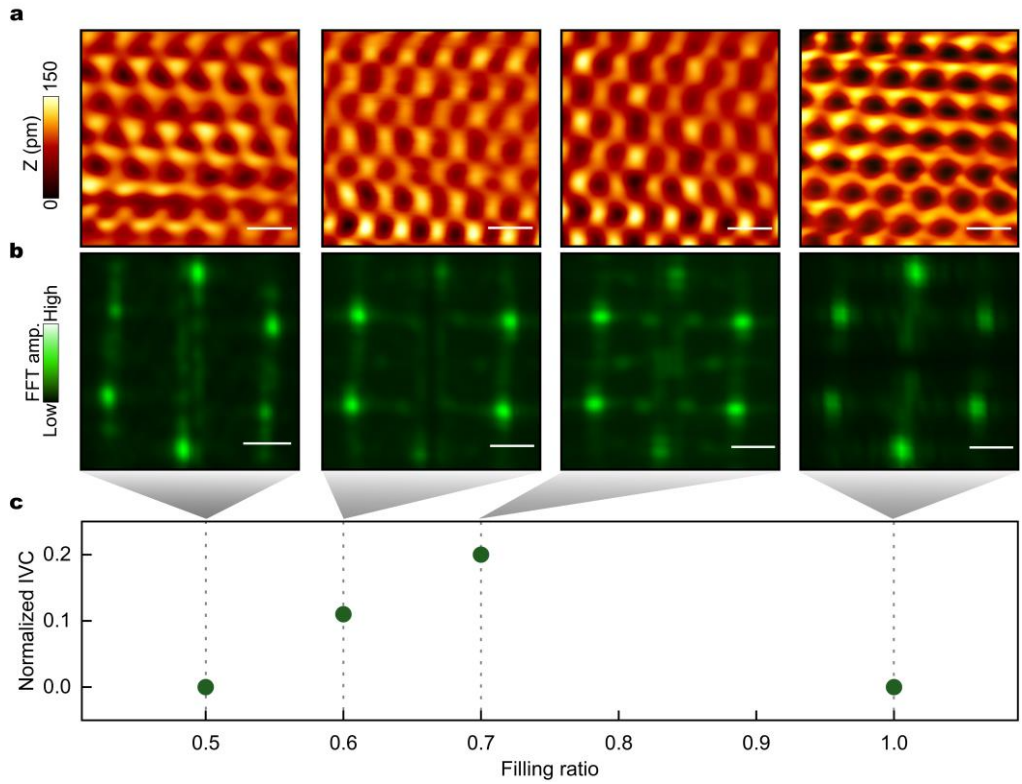

**Fig. 4 | Filling dependence of IVC patterns**. **a**, Atomic-resolution STM topographic images (1.5 nm × 1.5 nm, $V_{bias}$ = 40 mV, $I$ = 0.1 nA) of the tetralayer RG (device A) measured at different regions with varying filling ratios of the flat band. **b**, Corresponding FFT images of **a**. **c**, Normalized IVC strength, $\Sigma|\text{FFT(IVC)}/\Sigma \text{FFT(graphene)}|$, as a function of flat-band filling ratio. The IVC order is observed in ~60% and ~70% filled flat bands, while it is absent in nearly half-filled and full-filled flat bands. Scale bars: 3 Å (**a**); 2.5 nm$^{-1}$ (**b**).

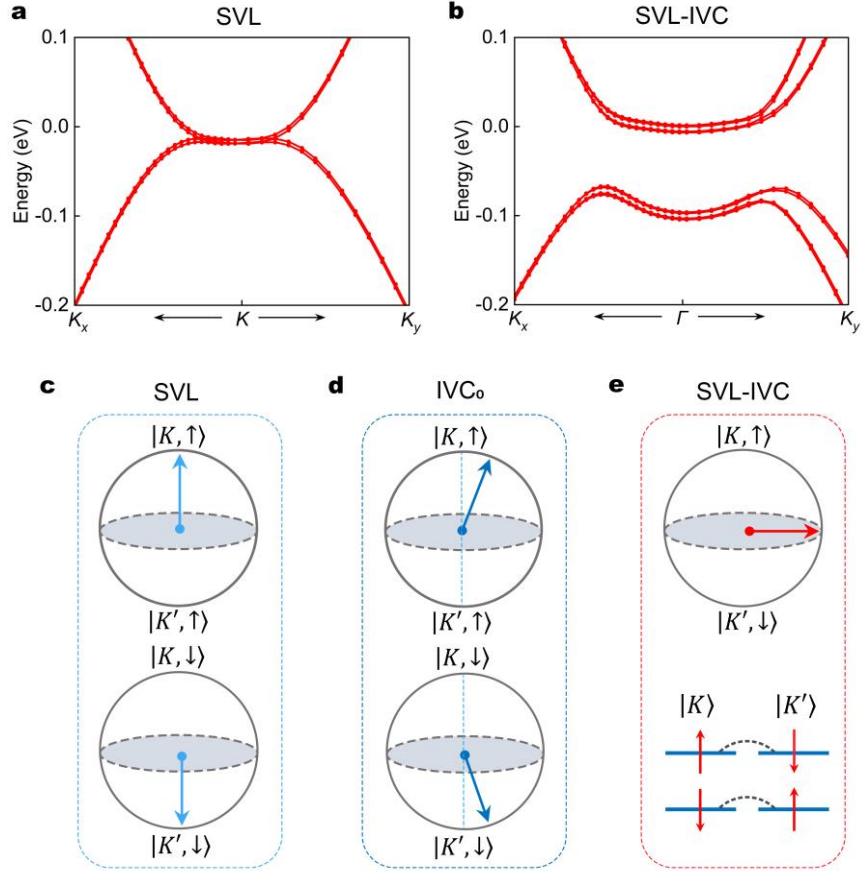

**Fig. 5 | Hartree-Fock calculations of IVC order. a**, The noninteracting bands of SVL states in the rhombohedral tetralayer graphene. Here Ising SOC $\lambda_I$ is set to 1 meV. **b**, The Hartree-Fock energy bands of SVL-IVC states at the doping electron concentration $0.9 \times 10^{12}$ cm$^{-2}$. **c-e**, The Bloch spheres of noninteracting SVL, IVC$_0$, and SVL-IVC states, respectively. The electron-electron interaction assisted intervalley spin-flip hopping are also shown in **e**.